\documentclass[nofootinbib,prl,amsmath,amssymb,twocolumn,floatfix,onecolumn]{revtex4-2}

\usepackage{tikz}
\usetikzlibrary{shadings}
\usepackage{graphicx}% Include figure files
\usepackage{dcolumn}% Align table columns on decimal point
\usepackage{bm}% bold math
\usepackage{amsmath}

\usepackage{lineno}
\begin{document}

\preprint{APS/123-QED}

\title{Acoustic Lattice Resonances and Generalised Rayleigh--Bloch Waves}

%\title{Generalised Rayleigh Bloch Waves Predict Acoustic Lattice Resonances}

\author{G.~J.~Chaplain$^{*,1}$, S.~C.~Hawkins$^{2}$, M.~A.~Peter$^{3}$, L.~G.~Bennetts$^{4}$ and T.~A.~Starkey${^1}$}

\affiliation{$^1$Centre for Metamaterial Research and Innovation, Department of Physics and Astronomy, University of Exeter, United Kingdom \\ 
$^2$ School of Mathematical and Physical Sciences, Macquarie University, Sydney, NSW, Australia \\
$^3$ Institute of Mathematics, University of Augsburg, Germany and\\ Centre for Advanced Analytics and Predictive Sciences (CAAPS), University of Augsburg, Germany \\
$^4$ School of Computer and Mathematical Sciences, University of Adelaide, Australia 
}

% Difficult to predict resonances. Connection with extended RB waves. They allow us to understand when these resonances occur. Depends on the aspect ratio. You need to be able to predict and design where the resonances are - there are goverened by the RB waves! 
%Careful with wording of phenomenon regarding scattering coefficient  

% \begin{abstract}
% We experimentally observe radiative acoustic lattice resonances along a diffraction grating and connect them to generalised Rayleigh--Bloch waves above the cut-off, considering both short and long arrays of non-resonant 2D cylindrical Neumann scatterers embedded in a dispersionless fluid (air). On short arrays, we observe finite lattice resonances under continuous wave excitation, and on long arrays, we observe propagating Rayleigh--Bloch waves under pulsed excitation. We consider multiple wave scattering theory and, in doing so, unify differing nomenclatures used to describe waves on infinite periodic and finite arrays and the interpretation of their dispersive properties.
% \end{abstract}

\maketitle

\noindent\textbf{Abstract}
The intrigue of waves on periodic lattices and gratings has resonated with physicists and mathematicians alike for decades. In-depth analysis has been devoted to the seemingly simplest array system: a one-dimensionally periodic lattice of two-dimensional scatterers embedded in a dispersionless medium governed by the Helmholtz equation. We investigate such a system and experimentally confirm the existence of a new class of generalised Rayleigh--Bloch waves that have been recently theorised to exist in classical wave regimes, without the need for resonant scatterers. Airborne acoustics serves as such a regime and here we experimentally observe the first generalised Rayleigh--Bloch waves above the first cut-off, i.e., in the radiative regime. We consider radiative acoustic lattice resonances along a diffraction grating and connect them to generalised Rayleigh--Bloch waves by considering both short and long arrays of non-resonant 2D cylindrical Neumann scatterers embedded in air. On short arrays, we observe finite lattice resonances under continuous wave excitation, and on long arrays, we observe propagating Rayleigh--Bloch waves under pulsed excitation. We interpret their existence by considering multiple wave scattering theory and, in doing so, unify differing nomenclatures used to describe waves on infinite periodic and finite arrays and the interpretation of their dispersive properties.

\smallskip
\noindent{\textbf{{Introduction}}
\smallskip
% \noindent The intrigue of waves on periodic lattices and gratings has resonated with physicists and mathematicians alike for decades. In-depth analysis has been devoted to the seemingly simplest array system: a one-dimensionally periodic lattice (grating) of two-dimensional scatterers embedded in a dispersionless medium governed by the Helmholtz equation.

One-dimensionally periodic lattices (with appropriate boundary conditions on the comprising scatterers \cite{wilcox1984scattering}) support surface waves that propagate along and exponentially decay away from the lattice. These exist under many guises for both scalar and vector wave regimes from acoustics and electromagnetism, to linear water waves and elasticity \cite{porter_evans_1999,peter_meylan_2007, antonakakis2014asymptotic,colquitt2015rayleigh}. Distinct from naturally occurring interfacial surface waves (e.g.~Rayleigh, Stonely, and Scholte waves, or surface plasmon polaritons), the waves we allude to owe their existence to the underlying embedded periodic lattice and are thus commonly termed Rayleigh--Bloch (RB) waves; they exist on infinite lattices, with the criterion for their existence having been analysed in the context of functional analysis for some time \cite{bonnet1994guided,porter_evans_1999,linton_mciver_2002}. In particular they differ from pure Bloch waves that exist in cells which are bounded (e.g.~an infinite 2D phononic crystal), as RB waves exist in 1D lattices with an unbounded unit cell, resembling a strip or ribbon \cite{CHAPLAIN2019162}.

The finite width of the periodic cells (or strips) tiling physical space gives rise to multiple `cut-offs': wavevectors in reciprocal space at the Brillouin Zone boundaries (BZBs) \cite{Brillouin1953}, with associated frequencies determined by the dispersion relation of the array. The wavelengths of the associated Bloch solutions become integer half-multiples of the cell width and standing wave solutions are formed (the Bragg condition). For a 1D periodic lattice embedded in a dispersionless medium, RB waves within the first BZ (below the first cut-off) are constrained to lie below the sound-line (in acoustics), the gradient of which gives the velocity of free-space acoustic waves. For symmetric scatterers, RB waves in the standing-wave limit in this region are equivalent to so-called Neumann modes \cite{porter_evans_1999}. In a finite array, % thecorresponding near-standing-wave solutions 
%nearby RB waves (almost standing waves) create 
they are closely related to localised lattice resonances, which have the form of near-standing-wave solutions just below the Neumann-mode frequency created by constructive interference between RB waves after end reflections \cite{thompson2008new}; the frequency at which they exist is often predicted by analysis of the infinite periodic picture. 

\begin{figure*}[b]
    \centering
    \includegraphics[width = 0.9\textwidth]{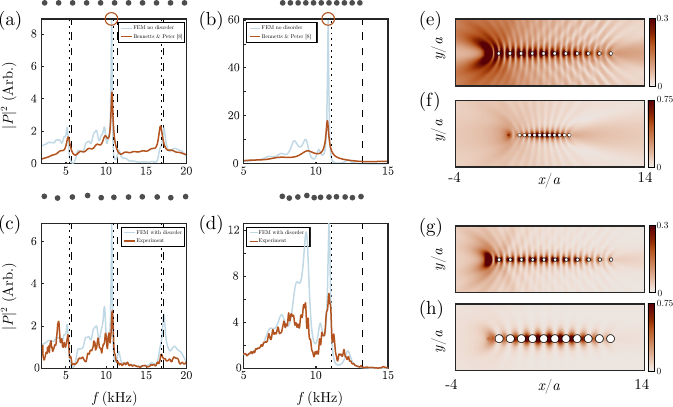}
    \caption{Acoustic Lattice Resonances. (a,b) Normalised load on central cylinder from FE (blue -- with point source) and method in \cite{bennetts_peter_2022} (orange -- with plane wave source), for an 11 scatterer array with $r/a = 0.15$, $0.35$. (c,d) Analogous FE curves (blue) on disordered array and experimental results (orange). The band edges and standing-wave frequencies from the infinitely periodic analogue are shown in dashed and dotted lines, respectively. Array schematics shown above panels. (e,f) Normalised absolute pressure field from FE method at frequencies marked with circle in (a,b) respectively. (g,h) Analogous multiple wave scattering simulations but with radii scaled rather than separation. Point source excitation is to the left of each array.}
    \label{fig:Resonances}
\end{figure*}

Notably, there has been more recent mathematical consideration of RB waves that exist above the first cut-off \cite{PORTER200529}. They exist in the radiative (leaky) regime, i.e.~above the sound line, and can therefore couple to plane waves. In particular, a second kind of RB mode exists above the first cut-off in which their wavenumbers become complex-valued. The connection between these ``generalised'' RB waves\footnote{At their discovery, they were termed extended RB waves \cite{bennetts_peter_2022} due to where they exist in terms of the position of the cut-off. Here we denote them generalised RB waves so to avoid confusion with their extent in decay away from the array.} and finite lattice resonances has been explored \cite{bennetts_peter_2022}. They exist over a range of parameter intervals that we explore here for the first time experimentally, viz.~in audible acoustics. The modes of open systems are commonly termed quasi-normal modes \cite{laude2023quasinormal}, with special cases being higher-order Neumann and Dirichlet modes (see below) with purely real point eigenvalues that exist for very particular parameter combinations, embedded in the continuous spectrum. As such, they are often termed bound states in the continuum (BIC) \cite{kang2023applications}. The generalised RB waves we consider here are not BICs, but very close to (sometimes referred to as quasi-BICs or leaky resonances
\cite{hsu2013observation,hsu2016bound}). They differ from embedded RB waves above the cut-off (i.e.~on infinite arrays \cite{PORTER200529}) and RB waves on semi-infinite gratings (that can be excited by plane waves \cite{peter_meylan_2007, linton2007scattering}) in that they exist \textit{both} above the cut-off and on finite arrays. 
% \textcolor{red}{Vincent Laude has something on this not out yet with penetrable scatterers - uses a nice method to find them like here \cite{laude2018stochastic}.}

% that are pressure waves in a fluid localised to a patterned rigid surface, opposed to surface acoustic waves such as Rayleigh waves \cite{brekhovskikh1959surface}. 

In this article, we seek to investigate the behaviour of lattices for acoustic waves in the regime above the first cut-off for short arrays with a few tens of scatterers (under continuous wave excitation) and long arrays with 
several tens of scatterers (under pulsed excitation), making the connection to generalised RB waves. We demonstrate that their existence depends on the aspect ratio between scatterer radius and cell width, as predicted in \cite{bennetts_peter_2022}, and justify this with multiple wave scattering (MWS) theory.

In Fig.~\ref{fig:Resonances}, we provide motivation for this study by showing experimental observations of acoustic lattice resonances above the first cut-off as predicted in \cite{bennetts_peter_2022}, for a finite array of 11 scatterers. The frequencies of the resonances are close to the BZBs (marked by dashed vertical lines).
Figures~\ref{fig:Resonances}(a,c) compare finite element (FE) models to the theoretical predictions of \cite{bennetts_peter_2022}, and FE models to experiment respectively (methods detailed below) for the case of radius to separation $r/a = 0.15$.  In Figures~\ref{fig:Resonances}(b,d), we show analogous results but for the case where $r/a = 0.35$ where no such higher resonances exist, as predicted \cite{bennetts_peter_2022}. The FE simulations in Figs.~\ref{fig:Resonances}(c,d) include a small amount of disorder in the centre position of the cylinders to reflect the experimental limitations (see below), which is known not to localise the RB wave \cite{bennetts2017localisation}. In Figs.~\ref{fig:Resonances}(e,f), we show FE frequency domain simulations of the normalised acoustic pressure, showing the lattice resonances, for the cases of $r/a = 0.15$ and $r/a = 0.35$ respectively. A point source excitation is to the left of the array, at the corresponding frequencies marked by the circles in (a,b). In these examples, the radius is fixed and the spacing is altered, as in the experiment. In Figs.~\ref{fig:Resonances}(g,h), we show multiple wave scattering simulations for the same configurations, but this time with the spacing remaining fixed and the radius changing. Coupling to radiation of the resonance above the first cut off is evident in (e,g).
% Throughout this paper we shall connect the existence of these modes to extended RB modes on infinite gratings, and provide context from multiple wave scattering theory for their existence.

Lattice resonances above the cut-off have been investigated for systems of \textit{resonant} scatterers for many years, particularly in the active field of plasmonics \cite{maradudin2014modern,kravets2018plasmonic,PhysRevB.74.205436,markel1993coupled,cherqui2019plasmonic} with leaky antennas serving as an attractive application \cite{PhysRevB.82.144305,Bulgakov:16}. However, the collective electromagnetic oscillations in plasmonics differ fundamentally from the acoustic lattice oscillations we observe here; in plasmonics the scatterers have their own individual resonant profiles described by well-defined dielectric functions, given by for example the Drude model \cite{bohren2008absorption}. We consider classical, linear acoustic pressure waves on a 1D periodic lattice of Neumann (impenetrable, sound-hard) cylinders at audible frequencies in air. 
% \textcolor{red}{might need to mention Vincent Laude's cases here?}
We experimentally observe collective acoustic lattice resonances above the (first) cut-off that arise due to radiative coupling of scattered waves by the grating. This is also viewed numerically with both the finite element method and semi-analytical multiple wave scattering theory, leading to a unification of nomenclature between the methods of analysis. Finally, we elucidate the physical mechanism driving these resonances, and how this manifests itself in their experimental detection by considering scatterer polarisability.

\smallskip
%\tableofcontents
\noindent{\textbf{{Acoustic Rayleigh--Bloch waves}}
\label{sec:ARB}
\smallskip
\begin{figure*}
    \centering
    \includegraphics[width = 0.95\textwidth]{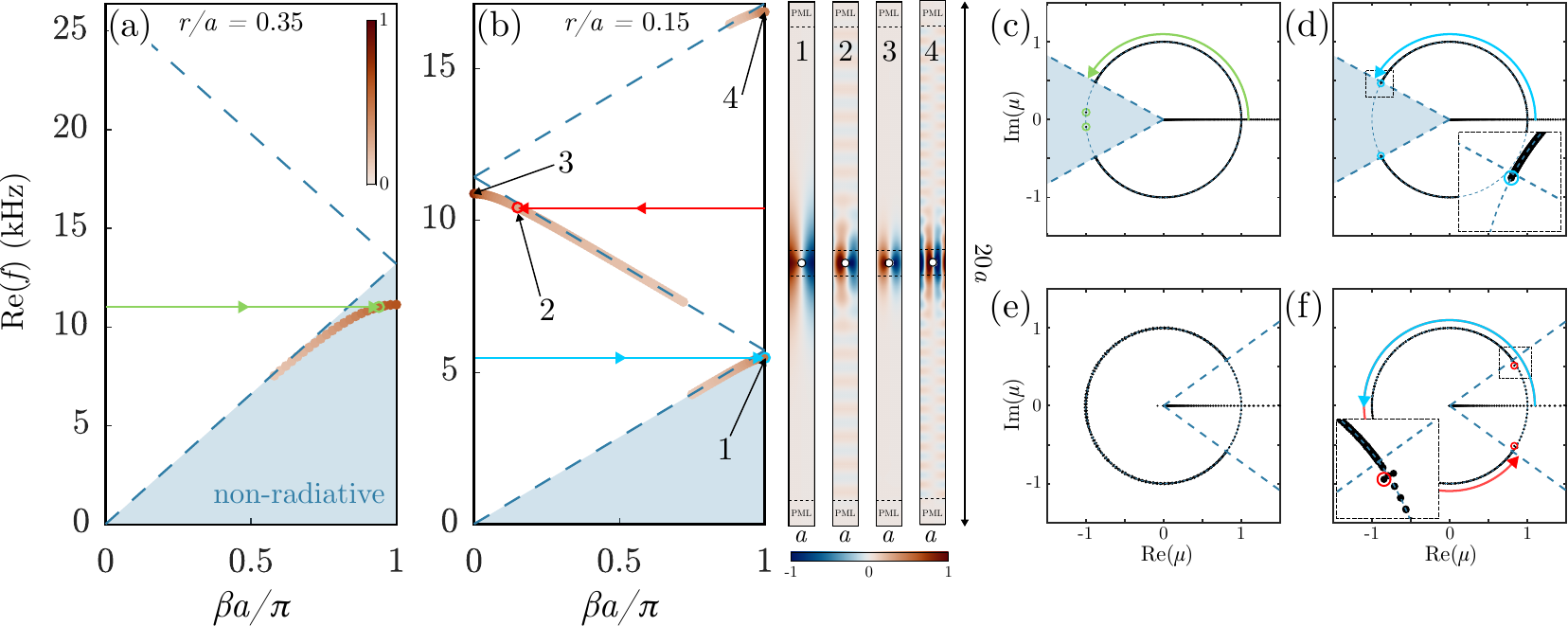}
    \caption{Dispersion curves and RB modes. (a,b) FE dispersion curves for differing aspect ratio $r/a = 0.35$, $0.15$, respectively. The sound line is shown in dashed blue with non-radiative regime highlighted. Colourscale indicates localisation to the array (Methods). Mode-shapes show RB and generalised RB modes numbered in (b). Neumann and Dirichlet trapped modes (purely real eigenfrequencies) are labelled 1\&3, respectively. (c-f) Representations of dispersion spectra as complex eigensolutions $\mu$ (black points) for $r/a$ = 0.35 (c,e) and 0.15 (d,f). Coloured arrows show direction around the continuous spectra along the unit circle, corresponding to a fixed frequency and altering $\beta$ in (a-b) -- the real parts of the eigensolutions are highlighted with corresponding colours in (a-b). Below the cut-off, the eigensolutions lead the unit circle, as in (c,d). The first generalised RB mode above the cut-off appears as complex solutions off the unit circle for large aspect ratio (f) that are not present in (e). Dashed boxes in (d,f) show a zoom near the cut-offs.}
    \label{fig:Dispersion}
\end{figure*}

% \begin{figure*}
%     \centering
%     \includegraphics[width = 0.75\textwidth]{FEM_disp_schem.pdf}
%     \caption{Dispersion curves and interpretations. (a-c) Dispersion curves obtained by the FEM shown in the Irreducible Zone scheme (on normalised in-plane wavevector) for differing aspect ratio $r/a = 0.15$, $0.25$, $0.35$ respectively. Colorscale indicates measure of degree of localisation to the array (see appendix). The sound line is shown in dashed blue with non-radiative regime highlighted below the first cut-off. (d-e) are representations of dispersion diagram as the complex solutions $\mu$ at the frequencies shown by the solid blue lines in (b): the arrows show the direction around the continuous spectra in (d-e) along the unit circle. (d) follows the lower blue arrows in (b) such that the mode is met below the cut-off and the eigensolutions lead the unit circle in the non-radiative regime. In (e) we near the second cut-off and the complex extended RB solutions appear off the unit circle, as in \cite{bennetts_peter_2022}, separate from the continuous spectrum. \textcolor{red}{get exact plots from Luke/Malte/Stuart of the case for above/below cut-off for r/a = 0.15.} \textcolor{blue}{Do we want to combine Fig 5 mode plots?}}
%     \label

% \textcolor{red}{fluctuations \cite{zeng2019fluctuation}
% \end{figure*}

\noindent To find acoustic RB waves, we seek solutions, $\phi$, of the 2D Helmholtz equation 
\begin{equation}
    (\Delta + k^2)\phi = 0,
    \label{eq:Hh}
\end{equation}
in the $xy$-plane with sound-hard (Neumann) boundary conditions applied on periodically spaced cylindrical scatterers with boundary $\Gamma$, separation $a$ and radius $r$, that satisfy both radiation and Bloch conditions, i.e. 
\begin{equation}
    \begin{cases}\frac{\partial\phi}{\partial \boldsymbol{n}} = 0 \quad \text{on $\Gamma$}, \\
    \phi \rightarrow 0 \quad \text{as $|y| \rightarrow \infty$}, \\
    \phi_{R} = \exp(iR\beta)\phi_{0}, \\
    \partial_{x}\phi_{R} = \exp(iR\beta)\partial_{x}\phi_{0},
\end{cases}
\label{eq:boundary}
\end{equation} 
where $\boldsymbol{n}$ is the outward surface normal, $\phi_{0}$ is the solution in the fundamental cell and $\phi_{R}$ is the solution in the $R^{\text{th}}$ cell at a distance $R = na$ for $n \in \mathbb{Z}$. Solutions to this problem are known as pure Rayleigh–-Bloch surface waves \cite{linton_mciver_2002} with the prescribed values of $k^2$ at which they occur being eigenvalues of the operator $-\Delta$ subject to \eqref{eq:boundary}. As is often the case, a non-dimensional form is adopted and, as such, $k$ is a wavenumber that acts as a proxy for a prescribed angular frequency. With this, $\beta$ represents the RB wavenumber, i.e.~the component of the wavevector parallel to the array. One is, of course, free to fix either $k$ or $\beta$ and solve for the other; differences in convention typically arising between mathematicians and physicists, e.g.~the methods in \cite{bennetts_peter_2022} and \cite{moore2023acoustic} respectively. Throughout, we shall use a mixture of both and highlight the advantages in finding and interpreting generalised RB waves.

We are careful to note here that the final two boundary conditions in \eqref{eq:boundary} represent the Floquet--Bloch conditions across the unit cell, satisfied only in the infinite problem. In the case of a finite array, near the cut-off frequencies the boundary conditions are often referred to as the Neumann and Dirichlet conditions \textit{across} the cell. The corresponding solutions being Neumann and Dirichlet modes respectively (labelled `1' and `3' respectively in Fig.~\ref{fig:Dispersion}). This nomenclature originates from trapped waves in a channel with such boundaries on the sidewalls \cite{linton_evans_1992, maniar_newman_1997,utsunomiya_taylor_1999,bennetts_peter_2022} and are also sometimes referred to as the periodic and anti-periodic conditions across the cell \cite{antonakakis2013high}. We make this point to avoid confusion; when Neumann (sound-hard) boundary conditions are referred to, it is on the boundary of the scatterers $\Gamma$ and not across the unit cell. Indeed, an infinite array of Dirichlet scatterers does not support RB waves \cite{wilcox1984scattering}.

In Fig.~\ref{fig:Dispersion}, we show two common interpretations of the dispersion relations for infinite arrays. The parallel component of the wavevector along the axis (the RB wavenumber $\beta$) is related to the free-space wavevector of magnitude $k_0 = \omega/c$, with $c$ the speed of sound through $\beta = k_{0}\sin\theta$, where $\theta$ prescribes the angle relative to the normal to the array axis. In Fig.~\ref{fig:Dispersion}(a,b), the dispersion curves for RB waves below and above the cut-off are evaluated with finite elements. Here, the RB waves below the first cut-off are highlighted in the non-radiative regime $\beta > k_0$ (shown in blue). They exist, with varying degrees of dispersion, for any aspect ratio $r/a$. Modes outwith the non-radiative regime, i.e.~above the first cut-off, are shown to exist below $r/a < 0.33$ as predicted by \cite{bennetts_peter_2022}. Mode shapes are shown in the unit-strip at four frequencies, labelled in Fig.~\ref{fig:Dispersion}(b). Modes 1\&3 are the Neumann and Dirichlet trapped modes respectively, with purely real eigenfrequencies. Modes 2\&4 represent generalised RB waves that have complex eigenfrequencies, evidenced by the coupling to the far-field. 

In Fig.~\ref{fig:Dispersion}(c-f), we adopt the representation from \cite{bennetts_peter_2022}, and show the RB waves through the introduction of the parameter $\mu$: the eigenvalues of the transfer operator that describes propagation along the array (method outlined in \cite{bennetts2017localisation}). For RB waves, the eigenspectra take the form $\mu = \exp(\pm i\beta)$ and RB modes below the cut-off arise as eigenvalues that lead the unit circle in the complex plane in the non-radiative regime (green points in Fig.~\ref{fig:Dispersion}(c)). Those that lie on the unit circle are solutions for propagating background waves forming the continuous spectra. This corresponds to following $\beta$ at a fixed frequency, highlighted by the green arrow in Fig.~\ref{fig:Dispersion}(a), reaching the highlighted eigenvalue in green. At the first cut-off, the unit circle closes at $\mu= -1$ ($\beta = \pi/a$). Above the first cut-off $\beta > \pi/a$ (now in a higher BZ we see as band folded in Fig.~\ref{fig:Dispersion}(b)), the continuous spectrum overlaps itself and generalised RB waves with complex wavenumber appear as pairs of eigenvalues off the unit circle, shown in Fig.~\ref{fig:Dispersion}(f). When the aspect ratio becomes large, i.e.~the cylinder occupies a larger fraction of the unit strip, these modes above the cut-off cease to exist \cite{Keipaper}; they are not present in (e) for the case of $r/a = 0.35$. In the following sections, we detail this observation experimentally and explain this phenomenon in terms of multiple wave scattering theory and the polarisability of the scatterers, in particular, matching the scattering amplitudes of the radiative wave functions with the symmetries of the eigenmodes supported by the array. We elucidate this for the first generalised RB mode above the cut-off that approaches the Dirichlet trapped mode \cite{PORTER200529}.

% \textcolor{red}{In Appendix A. we provide an overview of the detail in the numerical methods employed to solve \eqref{eq:Hh}-\eqref{eq:boundary} that comprise the FEM using COMSOL multiphysics \cite{comsol}, a semi-analytical model based on the spectra of the associated transfer operator of the array \cite{bennetts2011wave, bennetts2017localisation, bennetts_peter_2022}, and a reduced order model for multiple scattering theory via the transition matrix method (T-matrix) \cite{stuartTMAT,StuartTMAT2}. Of course one can utilise other numerical methods such as finite difference or spectral collocation to solve the ensuing eigen-problem in the physical domain. The advantage of comparing all three methods is we are able to flexibly solve both the finite and infinite problems and compare nomenclature between the dispersion spectra outlined above.} 
\smallskip
\noindent\textbf{{Experimental Results}}
\smallskip
\begin{figure*}[t]
    \includegraphics[width = 0.9\textwidth]{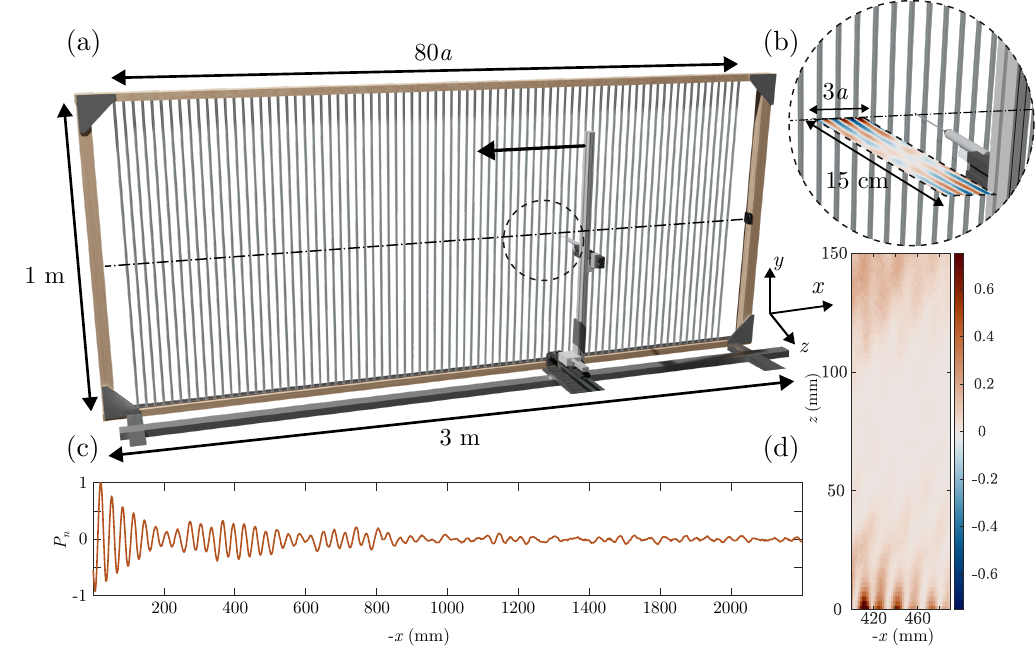}
\caption{Experimental setup and example scans. (a) Scanning $xyz$-stage and long array of 80 cylindrical acrylic rods in laser-cut stencils that set $r/a$. Dash-dotted line shows scan path and arrow shows direction in $-x$. (b) Zoom of area scan and example spatial distribution showing normalised real pressure amplitude at a frequency of $10.9$ kHz, for $r/a = 0.15$. (c) Normalised real pressure Fourier amplitude at $10.9$ kHz evaluated along scan line. (d) Corresponding time-averaged absolute pressure field area scan, highlighting localisation to the array; the decay due to radiative loss evident in (c-d) with beating in amplitude occurring due to interference with source (that decays with distance as expected).}
\label{fig:setup}
\end{figure*}

% \textcolor{red}{Is it possible to highlight in the first paragraph which is the ``main'' experimental result
% and maybe summarise it. Some of the experimental results are in Fig.~1 but the main purpose of Fig.~1 is to 
% motivate the rest of the paper. Would it be worth pulling those out of Fig.~1 and putting them in this section too?}
\noindent To observe the generalised RB waves above the cut-off experimentally, we consider two arrays with aspect ratios $r/a$ = 0.15 and 0.35, fabricated by fixing acrylic rods ($r = 4.5$\,mm) in slotted laser-cut templates mounted within a timber frame. We design two array lengths: a set of 11 rods, and a much larger array of 80 rods to approximate the infinite regime and to achieve the required resolution in reciprocal space. The rods are 1\,m in height and a loudspeaker was mounted within the frame, at the rod mid-height. Measurements are extracted from the mid-plane to approximate an open 2D array. A schematic of the array and scanning equipment is shown in Fig.~\ref{fig:setup}(a) and detailed experimental procedures are outlined in the Methods section.

In Fig.~\ref{fig:Resonances}, we show the results for the case of 11 cylinders under continuous wave excitation. Figures~\ref{fig:Resonances}(a,b) show the normalised load on the central cylinder, evaluated through frequency domain FE simulations, and the method outlined in \cite{bennetts_peter_2022}. Clear peaks can be seen that correspond to collective lattice resonances with frequencies close to the cut-offs (i.e.~the band edges of the infinitely periodic case, shown by the dashed vertical lines). Also shown are the predicted standing-wave solutions from the corresponding infinite array (e.g.~the frequency of modes 1, 3 etc., extracted from the dispersion curves in Fig.~\ref{fig:Dispersion}(b)) as dotted vertical lines. 

In practice, the rods are nonuniformly bowed in the manufacturing procedure. To explore the impact of this,
in Fig.~\ref{fig:Resonances}(c,d), we show a similar comparison, but this time between FE and experiments, and in the FE case disorder has been added to the scatterer positions (in the form of normally distributed random noise). As expected, RB waves still exist in the case of disorder \cite{bennetts2017localisation}. The pressure amplitude is extracted at positions near the central rod with and without the array, averaged and normalised. There is good agreement between the predicted positions of the frequencies of the resonances, and their oscillatory behaviour before the cut-off \cite{zeng2019fluctuation}. They closely align to the eigensolutions for the standing waves in the infinitely periodic case, indicating the nature of their existence through reflections of generalised RB waves.

The RB waves are clearly observed in the longer array.
We perform pulsed measurements along an array with 80 rods to approximate an infinite array when excited with an acoustic pulse. To obtain the dispersion spectra, the microphone is scanned along the full sample length in the propagation direction. To visualise the wave-field, small area pressure fields maps are shown on a plane normal to the rods in the propagation direction (150\,mm x 3 unit cells) in Fig.~\ref{fig:setup}(b). A zoom of the absolute time-averaged pressure field highlights localisation to the array. In Fig.~\ref{fig:setup}(c), we show the normalised real pressure frequency response near the second cut-off, at 10.9\,kHz, showing propagation of a decaying generalised RB wave. Due to the intrinsic losses of the fluid and the leaky coupling to radiation, the amplitude decays along the array and as such reflections from the end are minimal; the resonances observed in Fig.~\ref{fig:Resonances} are not observed due to the pulsed excitation as steady state is not reached. 

In Fig.~\ref{fig:exp}, we show the frequency spectra obtained by Fourier analysis (see Methods). Dispersion spectra are normalized by the maximum value at each frequency for each aspect ratio. Fig.~\ref{fig:exp}(a,c) and insets show the logarithmic spectra to aid visualisation of the dispersion. In the case of high aspect ratio, as predicted, generalised RB waves do not exist and are only detected when the spacing between the cylinders is sufficiently large. In Fig.~\ref{fig:exp}(b,d) we show the numerical counterpart to the experiment, the results of similar Fourier analysis of a 2D frequency domain simulation using FE. The array is excited by a point source at the same distance away from the array to the loudspeaker. In this case, the zero group velocity mode at the band edge is detected as viscous losses are not included. Overlaid in both cases are the corresponding eigenfrequencies from the infinite dispersion problem (Fig.~\ref{fig:Dispersion}) showing excellent agreement.  

\begin{figure*}[t]
    \includegraphics[width = 0.95\textwidth]{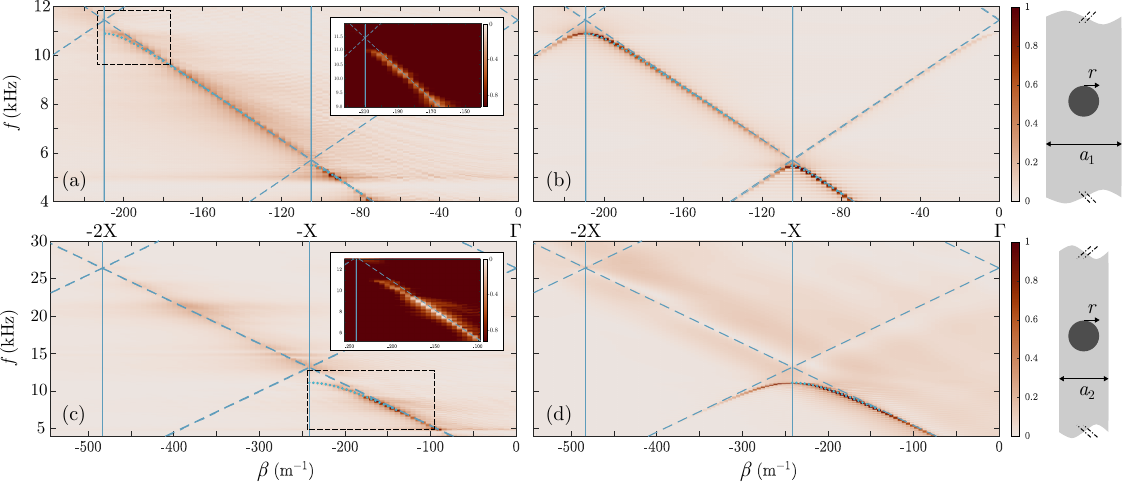}
\caption{Experimental and numerical observation. (a,b) Normalised experimental and numerical Fourier spectra respectively from line scans for the configuration of 80 rods with $r/a_{1} = 0.15$. The mode above the cut-off is predominant, as predicted in Figs.~\ref{fig:Resonances}(c)\&~\ref{fig:Dispersion}(b). (c,d) Show similar spectra but for the ratio $r/a_{2} = 0.35$ --- no mode above the cut-off is supported. Insets show logarithm of the Fourier spectra in the regions highlighted by the dashed rectangles, where the peeling of the mode off the sound-line is more visible. Overlaid points are from the eigenfrequency study (Fig.~\ref{fig:Dispersion}). Schematics show the unit strips.}
\label{fig:exp}
\end{figure*}

\smallskip
\noindent\textbf{{Discussion}}
\smallskip

\noindent Here, we outline a physical reasoning for why generalised RB waves do not exist at high aspect ratios. To do so, we turn to language that is commonplace within electromagnetism and plasmonics, and now the metamaterial community in general, and focus on the polarisability of the scatterers.

Electromagnetic polarisability is a measure of how easily the charge distribution of a scatterer, be it a molecule or metallic nano-particle, may be distorted by an external electric field \cite{barnes2016particle} and is thus related to the scattering strength of the material. In acoustics, an analogous notion of polarisability exists that relates the dipole and monopole scattering strengths (moments) to the particle velocity and local pressure \cite{sieck2017origins}. Measuring the tensor that governs these interactions has received attention in acoustics where the scatterers are considered acoustically small (i.e.~$ka \ll 1$) \cite{jordaan2018measuring,melnikov2019acoustic}. This is particularly relevant for sub-wavelength metamaterials but is not the case for the generalised RB waves, where, by definition, they have wavelengths commensurate with (and smaller than) the unit strip width. As such, we turn to multiple wave scattering (MWS) theory to elucidate the scattering profile of a single scatterer, then a pair of scatterers, and thereby inductively inferring the behaviour of the grating composed of many scatterers; this is commonplace in plasmonics, where the polarisabilty of a single metallic particle is used to obtain the scattering cross-section of an array \cite{hicks2005controlling,rodriguez2012collective}. We do so using the T-matrix method \cite{stuartTMAT}, which has had recent attention in metamaterial applications \cite{Hawkins2024}, to evaluate the radiating wavefunction expansion coefficients of the scattered field.  

Recall \eqref{eq:Hh}, the Helmholtz equation for a scalar monochromatic field (assuming $\exp(-i\omega t)$ time dependence). In MWS theory, the field $\phi$ is split into an incident field $\phi^{\text{inc}}$ that interacts with scatterers $\Gamma$, to induce a scattered field $\phi^{\text{scat}}$, both satisfying \eqref{eq:Hh} such that $\phi = \phi^{\text{inc}} + \phi^{\text{scat}}$. The scattered field must also satisfy, for an unbounded domain as we assume here, the Sommerfeld radiation condition \cite{colton1992inverse} 
\begin{equation}
    \lim_{\rho\rightarrow\infty} \sqrt{\rho} \left(\frac{\partial\phi^{\text{scat}}}{\partial \rho} -ik\phi^{\text{scat}} \right) = 0,
\end{equation}
which, in 2D, has the far field 
\begin{equation}
    \phi^{\infty}(\theta) = \lim_{\rho \rightarrow\infty}\sqrt{\rho}e^{-ik\rho}\phi^{\text{scat}}(\boldsymbol{r},\theta),
\end{equation}
where $\rho = |\boldsymbol{r}|$.
Solving for the field $\phi^{\text{scat}}$ then requires suitable boundary conditions to be imposed on $\Gamma$. We turn  to expansions in regular and radiating wavefunctions \cite{stuartTMAT}
\begin{align}
    \begin{split}
        \hat{\phi}_{n}(\boldsymbol{r},\theta)  &= J_{|n|}(k\rho)e^{in\theta}, \\ 
        \phi_{n}(\boldsymbol{r},\theta) &= H^{(1)}_{|n|}(k\rho)e^{in\theta},
    \end{split}
\end{align}
with $n \in \mathbb{N}$. $J_n$ and $H^{(1)}_n$ are the first-kind Bessel and Hankel functions of order $n$ respectively, such that 
\begin{align}
    \begin{split}
        \phi^{\text{inc}} &= \sum_{n}f_n \hat{\phi}_n , \\
        \phi^{\text{scat}} &= \sum_{n}A_n\phi_{n},
    \end{split}
\end{align}
where $f_n$ and $A_n$ are expansion coefficients,
and $f_n$ are typically known, and are known analytically for common incident waves such as plane waves. Numerically evaluating these requires truncation in the summation; the T-matrix method enables numerically stable computation of these coefficients \cite{StuartTMAT2}, ultimately providing insight into the coupling strength between scatterers.

\begin{figure}
    \centering
    \includegraphics[width = 0.45\textwidth]{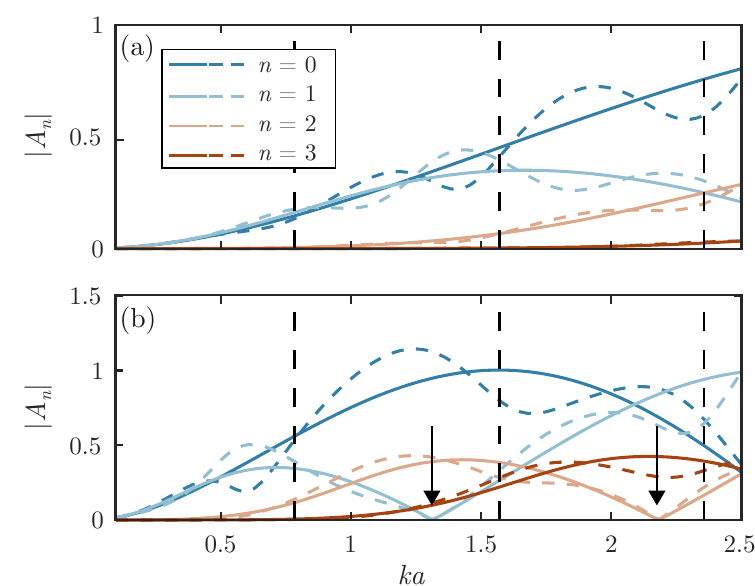}
    \caption{Radiating wavefunction expansion coefficients. (a) Coefficients of a single scatterer (solid lines) and for the first of a pair of scatterers (dashed) for fixed $r/a = 0.15$. (b) Corresponding curves for $r/a = 0.35$. Cut-offs marked by vertical dashed lines (increasing $n$ left to right).}
    \label{fig:scats}
\end{figure}
In Fig.~\ref{fig:scats}, we plot the radiating wave function expansion coefficients of (i) the scattered field of a single cylindrical scatterer (solid lines) and (ii) the coefficients for the first scatterer in a system of two scatterers separated by $a$ (dashed lines). In Fig.~\ref{fig:scats}(a), we show the evolution of the expansion coefficient amplitude as we vary $ka$ for fixed radius, corresponding to $r/a = 0.15$. For both the single and pair of scatterers, the expansion coefficients $|A_{n}|$ are non-zero near the $n^{\text{th}}$ cut-offs of the infinite regime (marked by dashed vertical lines). Considering the first generalised RB wave, the scatterers are sufficiently separated so that the dipole-like field can exist between the scatterers and they hence couple. Near the second cut-off, the amplitude of $|A_1| > |A_0|$ for the pair of scatterers, which agrees with the observations in Figures~\ref{fig:Resonances}(c),~\ref{fig:Dispersion}(b)\&~\ref{fig:exp}(a) that show this is the predominant mode. Examining the behaviour of the scattering coefficients, as we increase the number of particles in the array, is analogous to considering the increased scattering cross section of arrays of e.g.~resonant particles in plasmonics \cite{rodriguez2012collective}.

Contrasting this to Fig.~\ref{fig:scats}(b), where $r/a = 0.35$, we see that near the $n^{\text{th}}$ cut-off frequencies the corresponding $|A_{n}|$ vanishes (marked by black arrows) for the case of a single scatterer \textit{and} in the case of a pair of scatterers. The first scatterer cannot couple to the next with these radiating moments. By extension, considering many scatterers, a generalised RB wave cannot exist; the overlap integral (inner product) of the scattered field and that of the eigenmodes of the array vanishes and the modes are not supported. 

As an example, consider the case of the first generalised RB wave above the cut-off, which approaches the Dirichlet trapped mode ($\beta \equiv 0$ \cite{PORTER200529}), i.e.~mode 3 in Fig.~\ref{fig:Dispersion}. The dominant symmetry requires a dipole-like field between the scatterers; this is not supported when $|A_1| = 0$. At higher frequencies, $|A_1|$ is non-zero once more, but the next generalised RB wave requires matching to the quadrupole moment, but again we see near the cut-off $A_2 = 0$. For this reason, at large aspect ratios (e.g.~Fig.~\ref{fig:Resonances}(d) \& Fig.~\ref{fig:Dispersion}(a)) we do not see generalised RB waves. One could also consider asymptotic analysis of the gap as it becomes narrow for large aspect ratios \cite{vanel2017asymptotic,vanel2019asymptotic}.

\smallskip
\noindent \textbf{Conclusion}
\smallskip

\noindent Analysis of finite gratings is a well-trodden path, and their behaviour is often inferred from the infinitely periodic case, even with a very low number of repeat periods. This is commonplace across many wave regimes and has had success in antenna engineering \cite{skigin1999superdirective,skigin2009transmission}, with the validity of the periodic approach still receiving attention in the design of metamaterials 
\cite{sugino2017general,langfeldt2024validity}. Waves on finite and semi-infinite lattices are often labelled as RB waves, as is the case for electromagnetic waves on gratings \cite{barlow1954experimental,hurd1954propagation}; spoof surface plasmons \cite{pendry2004mimicking,hibbins2005experimental}; water waves with depth dependence \cite{peter_meylan_2007}; loaded thin elastic plates and in-plane elastic voids \cite{evans2007penetration,CHAPLAIN2019162,colquitt2015rayleigh}; or acoustic surface waves \cite{brekhovskikh1959surface,Kelders1998,moore2023acoustic}.

Here we have experimentally observed generalised RB waves above the first cut-off that manifest as acoustic lattice resonances confined to a diffraction grating. We considered 2D arrays of cylindrical Neumann scatterers embedded in air, forming both small and large-scale acoustic gratings for airborne sound in the audible frequency range. The first generalised RB mode was shown to exist over a range of frequencies for an example aspect ratio predicted by \cite{bennetts_peter_2022}. Generalised RB modes do not exist when the aspect ratio becomes large and the physical justification of this was presented in the context of multiple wave scattering theory; the scatterers do not support the radiating scattering wavefunctions which couple dipolar (and higher order) interactions. As we operate above the first cut-off, the supported waves couple through radiation and are inherently lossy.

In airborne acoustics, we are afforded the luxury of being able to achieve both continuous wave and pulsed excitations that permit the excitation of collective acoustic lattice resonances on short arrays and propagating generalised RB waves on long arrays respectively. In doing so, we have demonstrated that they are one and the same and that their interpretation is unified through the analogue of polarisability of the scatterers and the gaps between them.

\medskip
\noindent \textbf{Methods}
\medskip

\noindent\textbf{Finite Element Methods:}
Dispersion curves in Fig.~\ref{fig:Dispersion} are evaluated by an eigenfrequency study using the Acoustics module in COMSOL Multiphysics \cite{comsol}. Floquet--Bloch boundaries are added on the left and right sides of the modeshape geometries in Fig.~\ref{fig:Dispersion}, with perfectly matched layers (PML) top and bottom. The modes above the cut-off are isolated by evaluating the ratio of the integral of the absolute modulus of the pressure field in a region near the scatterer (dashed boxes in Fig.~\ref{fig:Dispersion}) to the same quantity in the rest of the domain \cite{chaplain2021metasurfaces}.

\smallskip

\noindent\textbf{{Continuous wave measurements:}}
Signal recorded on an oscilloscope (Siglent  SDS2352X-E). Acoustic data was recorded with a sampling frequency of 500 kSa s$^{-1}$, for a total time of 0.28 sec, and with 32 averages.

The experimental procedure was as follows: the microphone was positioned close to the central rod in the 11-cylinder long array. The lattice was driven to a steady state at discrete frequencies (2 to 20 kHz in 50 Hz steps) and the signal from the microphone recorded. This was repeated at three positions around the central rod and the data averaged. Similar measurements were performed at the same positions without rods for normalisation.

To determine the frequency response of the finite array, the temporal acoustic signals were summed, producing a signal (voltage) as a function of time $V(x_{i},t)$ at discrete positions $x_i$ near the central rod. The data were processed using temporal Fourier transform.

\smallskip
\noindent\textbf{Pulsed measurements:} The 80-rod long samples are excited by a tweeter (Kemo L010 Piezo Loudspeaker) mounted within the supporting frame at the mid-height of the rods. The loudspeaker is driven by an arbitrary waveform generator (Keysight 33500B), producing single-cycle Sine--Gaussian pulses centred at $f_\text{c} = 16$ kHz, and a broadband amplifier (Thurly Thandar Instruments WA301). The acoustic pressure field is measured with a small aperture microphone (Br\"{u}el {\&} Kjær Probe Type 4182 near-field microphone, with a preconditioning amplifier) positioned 2 mm normal to the array direction. Acoustic data are recorded by an oscilloscope (Picoscope 5000a) at sampling frequency $f_s = 312.5$ kHz. The microphone is mounted on a motorized \textit{xyz} scanning stage (in-house with Aerotech controllers), to map the acoustic signal spatially. An average was taken over 20 measurements at each spatial position to improve the signal-to-noise ratio. The microphone is scanned along the full sample length with 15 points per unit cell step-size in the propagation direction. Acoustic data are analysed using Fourier techniques to obtain the wavenumber--frequency dependence of the propagating waves. The fast-Fourier Transform (FFT, operator $\mathcal{F}$) of the measured signal voltage $V(x,t)$ returns the complex Fourier amplitude in terms of the wavenumber parallel to the surface $\beta$ and frequency $f$, $\mathcal{F}_x(|\mathcal{F}_t(V(x,t))|)$. Raw data are presented in Fig.~\ref{fig:exp} with no windowing or zero-padding in either space or time. 

\medskip
\noindent\textbf{Acknowledgements}
\smallskip

\noindent The authors would like to thank the Isaac Newton Institute for Mathematical Sciences, Cambridge, for support and hospitality during the program {\em Multiple Wave Scattering} supported by EPSRC grant no EP/R014604/1. T.A.S. acknowledges the financial support of Defence Science and Technology Laboratory (Dstl) through grants No. DSTLXR1000154754 and No. AGR 0117701). G.J.C. gratefully acknowledges financial support from the Royal Commission for the Exhibition of 1851 in the form of a Research Fellowship. The authors are also grateful to Dr. I.R. Hooper, Prof. S.A.R. Horsley and Prof. A.P. Hibbins for useful conversations.
S.C.H. gratefully acknowledges support from the Australian Research Council (ARC) grant no.~DP220102243.
L.G.B. also acknowledges ARC support (FT190100404). For the purpose of open access, the author has applied a ‘Creative Commons Attribution (CC BY) licence to any Author Accepted Manuscript version arising from this submission.
%TC:endignore

\smallskip
\noindent\textbf{{Data and Code availability}} 
\smallskip

\noindent The data and code that supports the findings of this study are available from the corresponding author upon reasonable request.

\smallskip
\noindent\textbf{Author Contributions}
\smallskip

\noindent MAP and LGB developed the theory of generalised RB waves. GJC and TAS performed FE simulations and carried out the experiments, and dealt with post-processing. SCH performed the MWS simulations and analysis. All authors contributed to the interpretation of results and writing of the manuscript.

% \smallskip
\noindent\textbf{Competing Interests}
% \smallskip

\noindent The authors declare no competing interests.

% \bibliography{apssamp}% Produces the bibliography via BibTeX.

%apsrev4-2.bst 2019-01-14 (MD) hand-edited version of apsrev4-1.bst
%Control: key (0)
%Control: author (8) initials jnrlst
%Control: editor formatted (1) identically to author
%Control: production of article title (0) allowed
%Control: page (0) single
%Control: year (1) truncated
%Control: production of eprint (0) enabled
\providecommand{\noopsort}[1]{}\providecommand{\singleletter}[1]{#1}%

\end{document}